# Enhancement of torque transmission capability in Magneto-Rheological fluid-based Clutch using novel hybrid corrugated plane transmission surface strategy

Jithin Vijaykumar
*Department of Mechanical Engineering*
*National Institute of Technology Calicut*
Calicut, India
vijayakumarjithin@gmail.com

Loyad Joseph Losan
*Department of Mechanical Engineering*
*National Institute of Technology Calicut*
Calicut, India
loyadjosephlosan@gmail.com

Saddala Reddy Tharun
*Department of Mechanical Engineering*
*National Institute of Technology Calicut*
Calicut, India
sreddytharun@gmail.com

Murthi Ram Chandra Reddy
*Department of Mechanical Engineering*
*National Institute of Technology Calicut*
Calicut, India
murthiramchandrareddy@gmail.com

Mood Rahul
*Department of Mechanical Engineering*
*National Institute of Technology Calicut*
Calicut, India
rc64998@gmail.com

Jagadeesha T
*Department of Mechanical Engineering*
*National Institute of Technology Calicut*
Calicut, India
jagadishsg@nitc.ac.in

*Abstract*— In an increased automated world, miniaturization is the key to widespread deployment of advanced technologies. Enhancing the torque transmissibility by abiding to the spatial constraints imposed by radial space availability has consistently remained a hurdle in the implementation of Magneto-Rheological (MR) clutches that use shear mode of MR fluid (MRF). This proves the necessity of a novel design capable of providing required transmission capability with a reduced transmission surface area. The present study analyzes a corrugated transmissible surface design which improves torque transmissibility with the help of increased transmission area and proper alignment of field lines passing through the MRF gap. In this paper, the impact of various dimensional parameters of a hybrid corrugated plane type MR clutch (MRC) design was studied with the aid of magnetic analysis performed on COMSOL Multiphysics software. The results obtained shows that various parameters in the design of MR clutches, such as annular and radial MR gaps, disc width, individual corrugation heights, corrugation width, bobbin thickness and radii of plane surface influences the torque transmission capability of MR clutches. Also, an optimization of the hybrid corrugated plane MR Clutch of the chosen geometry has been conducted with the transmission capability increasing by 39.37% compared with the non-optimized geometrical configuration.

## I. INTRODUCTION

Magneto-Rheological (MR) fluid is a smart fluid whose rheological properties can be precisely as well as instantaneously be manipulated with the variation in magnetic flux density, which in turn depends on the current value supplied to the electromagnets thus influencing the MRF. They are made up of small magnetizable particles suspended in a carrier fluid, usually silicone oil. MRFs have many applications in the field of engineering, such as in dampers, valves, clutches, and brakes [1-4].

Guoliang Hu et al. has proposed a MR brake design with double disc configuration considering proper material allocation to various components, achieving improvement in braking torque [5]. Thakur et al. studied on the various groove profiles added on the disc surfaces of MR clutches and experimentally proved their contribution in the enhancement of torque transmissibility [6]. Karthik et al. performed magnetic analysis to study the impact of MRF gap and disc width on braking torque [7]. To show the efficiency of the suggested method, Li et al. devised an optimization method for constructing MR clutches and carried out a comparison study between drum, single disc, and multi disc configurations of MR clutches to show the effectiveness of the method proposed [8]. Pisetskiy et al. experimented on the usage of permanent magnets to partially magnetize the MR clutch. It was found that a 50% reduction in MRF gap size resulted in an 85% improvement in torque to mass ratio [8]. Lee et al. showed that the clutch geometry and magnetic flux density are the parameters that influence the torque transmissibility of the MRC [9]. Wang et al. examined the thermal properties of a liquid-cooled MR clutch and presented a full design. It was discovered that the overall output torque decreased by 23.1% as temperature increased by 64.36 percent [10]. Dai et al. used experimental testing to determine that the torque transmission in their suggested design for the MR clutch, could be increased to 1.4 times that of the standard disc configuration clutch with equivalent dimensions [11].

## II. DESIGN AND MODELLING

The MR clutch comprises of the components namely:

1. Input (or driving) shaft.
2. Output (or driven) shaft.
3. Electromagnets formed by copper coils.
4. MR fluid

The topography of the hybrid disc type MR clutch studied is presented in Fig. 1. The shaft is made of structural steel (non-ferro magnetic) and the housing, along with the hybrid disc which is the combination of both plane and corrugated configurations, are made of low carbon steel which is highly permeable. The gap between the housing and the disc is occupied by the MRF.

In the proposed work, the width of corrugation and the separation of the corrugations are optimized so that maximal amount of torque can be transmitted by the MR clutch with respect to the coil placement.

The maximal torque that an MR clutch can transmit depends upon the shear strength of the MR fluid which in turn depends on the nature of MR fluid being used and the strength of the magnetic field applied. For the torque to be generated in shear mode, it is necessary that the magnetic field lines pass through the MR gap perpendicularly and hence the coils are placed near to the corrugations to allow maximum magnetic field lines to pass through the corrugated transmission surfaces.

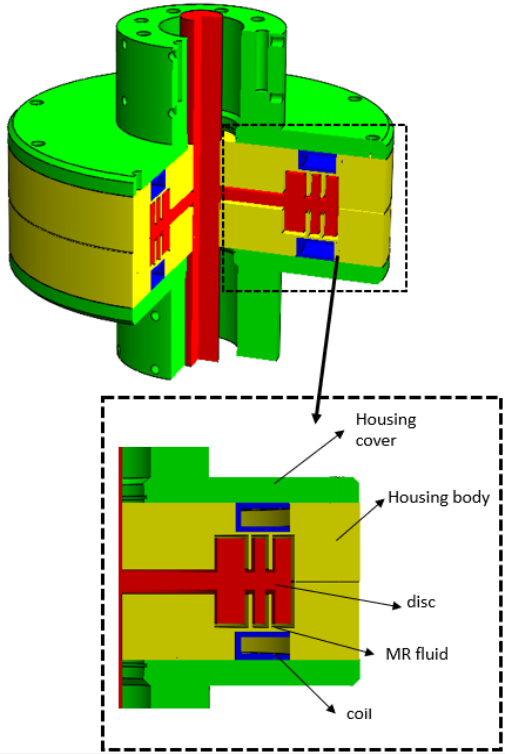

Fig. 1. 3-D CAD model for the hybrid corrugated MR Clutch

### III. MATHEMATICAL MODELING

The torque transmitted by the MR Clutch upon actuation by the current can be determined by using the basic torque equation given in equation 2.

$$dT = 2\pi r^2 \tau_z dr + 2\pi r^2 \tau_r dz \quad (2)$$

where r is the radial distance from the axis, $\tau_z$ and $\tau_r$ are the shear stresses along with axial and radial directions. The shear stress behavior of the MR fluid can be modeled using the Bingham-plastic model.

The term $2\pi r^2 \tau_z dr$ give the torque from a radial surface while the term $2\pi r^2 \tau_r dz$ gives the torque from an axial surface. $T_1$, the torque contributed by a single radial surface is given by:

$$T_1 = 2\pi \int_{r_1}^{r_2} r^2 \tau_z dr \quad (3)$$

$T_2$, the torque contributed by a single axial surface is given by:

$$T_2 = 2\pi r^2 \int_{z_1}^{z_2} \tau_r dz \quad (4)$$

The total output torque of the clutch can be found by adding together all the individual torques from all the radial and axial surfaces.

$$T = \sum_{i=1}^{n} T_1 + \sum_{j=1}^{m} T_2$$

where, n is the number of radial surfaces and m is the number of axial surfaces in the axisymmetric model of the MR Clutch.

### IV. MAGNETIC ANALYSIS OF INDIVIDUAL PARAMETRIC STUDY

The magnetic analysis for the MR clutch model was performed using COMSOL Multiphysics software using Finite Element Method. axisymmetric analysis was done exploiting the symmetric nature of the model, where Maxwell's equations were solved by the solver using a non-linear method based on Newton-Raphson method. The element type employed was a triangular element due to its easy configurability to the complex geometry of the MR Clutch.

The major design parameters that influence the torque transmission capability of the clutch were identified as corrugation height, corrugation width, annular MR gap, radial MR gap, disc width and the distance at which the corrugations start. The values of these parameters are varied individually at first to study their effect on the torque transmitted by the clutch. Finally, these values were parametrically swept together using the optimization module of COMSOL Multiphysics software, where the penalty method was employed to find the most optimized clutch design in terms of torque transmission capability. The materials assigned in the analysis for the various parts of the MR clutch like the housing, the disc, the coil along with the respective saturation magnetic flux densities are provided in TABLE I.

TABLE I.  MATERIALS ASSIGNED AND MAGNETIC PROPERTIES OF VARIOUS COMPONENTS IN THE MR CLUTCH

| Component | Material | Relative Permeability | Saturation Flux Density |
|---|---|---|---|
| Magnetic core | Low Carbon Steel | B-H Curve | 1.55 T |
| Coil | Copper | 1 | - |
| MR fluid | AMT Magnaflow + | B-H Curve | 1.65 T |
| Bobbin and others | Structural Steel | 1 | - |

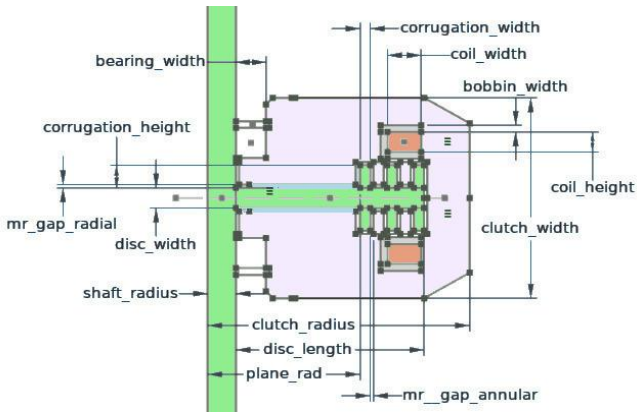

Fig. 2. Geometrical sketch and associated parameters used for the hybrid corrugated MR Clutch

TABLE II.  THE PARAMETERS EMPLOYED

| Parameter | Value |
|---|---|
| Clutch radius | 78 mm |
| Clutch width | 60 mm |
| Number of turns of coil | 600 |
| Current in the coil | 2.5 A |
| Radial surface MR Gap | d |
| Annular surface MR Gap | D |
| Radial distance of the first corrugation on the disc | plane_rad |
| Bobbin thickness | bobbin_width |
| Disc width | disc_width |
| Corrugation height | corrugation_height |

The magnetic analysis study and inferences drawn upon the torque transmission capability of the MR Clutch in terms of the effect of chosen variables, which are denoted in parametric form in Fig.2. are discussed below:

*A. Radial surface MR Gap (d) and annular surface MR Gap (D)*

The MR gap, which is the axial and annular gap between the disc and the housing, where the MR fluid resides is a critical design parameter that influences the torque transmission capability as this greatly influences the viscous component with the variation in shear rate. Secondly, the MR gap has the capability to configure the system to have a higher average magnetic field flux density, thereby influencing the magnetic field dependent yield stress. The surface plot of the torque value obtained for different combinations of Radial surface MR Gap(d) and Annular surface MR Gap (D) for the hybrid corrugated MR clutch under consideration is shown in Fig 3.

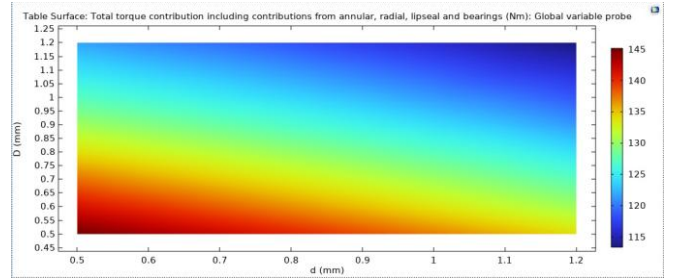

Fig. 3. Surface plot of the total transmitted torque with variations in the annular (D) and the radial gaps (d).

It is seen that the torque value is inversely proportional to both the MR gaps. It is because shear stress in the MR fluid is directly proportional to the velocity gradient according to Newton's law of viscosity and the velocity gradient is inversely proportional to the MR gap. Also, a lower MR gap ensures that the magnetic flux density is uniform throughout the transmission surfaces and yields better chain formation for the fibrillations.

Also, it is noticed that the parameters of radial and axial MR gaps don't cast equal influence over the torque transmission characteristic. Fig.3 suggests that it is more consequential to have a smaller annular gap than a smaller radial gap for the utility of higher transmission capability in the designed MR clutch.

## B. Bobbin Thickness (bobbin_width)

The bobbin is made of a comparatively less magnetically permeable material than the housing. This ensures that the bobbin redirects the magnetic field lines to ensure a longer flux route perpendicular to the MR fluid gaps. The thickness of the bobbin thereby determines the directionality of the magnetic flux reaching the MR fluid and hence the torque value of the clutch.

As the thickness of the bobbin increases the distance between the MR fluid and the coil also increases. When the bobbin thickness is increased and the coil is placed farther away from the MR fluid, and hence the torque value of the clutch decreases. The variation of torque value with respect to change in bobbin thickness is given in Fig 4.

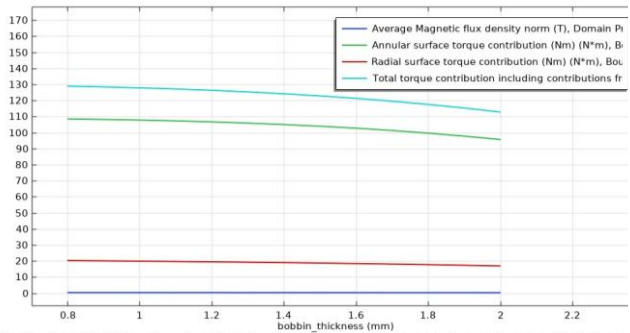

Fig. 4. 2D plot of the variation of the total torque transmitted with change in bobbin thickness.

## C. Disc Width and Corrugation Height (disc_width and corrugation height)

It was observed that in Fig.5 the average magnetic flux density increased in the MR fluid domain with the disc width being high and the corrugation height being small. But, when the total torque variation was plotted as in Fig.6, the higher average magnetic flux density didn't translate to higher transmission capability. The probable reason for this could be the lack of sufficient transmission surfaces occurring in the case of lower corrugation height.

With larger corrugation height, it was observed that the transmission area augmentation overtook the deficiency in the magnetic flux density and hence, the transmission capability increased. Also, in Fig.6., the existence of an optimum combination of disc width and corrugation height exists. This is achieved through a balance between the magnetic flux density and the transmission surface area.

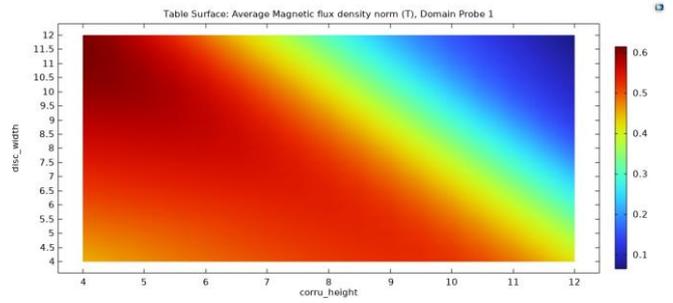

Fig. 5. Surface plot of the variation of average Magnetic flux density with respect to disc width and corrugation height.

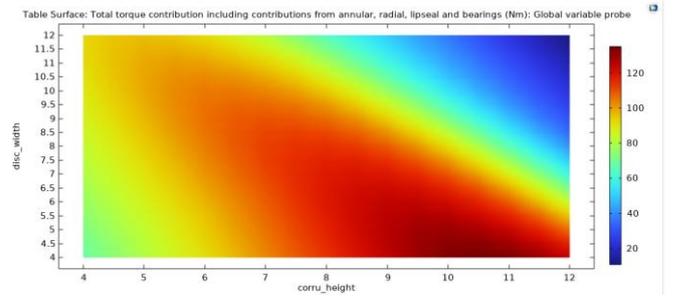

Fig. 6. Surface plot of the variation in total torque transmitted with respect to disc width and corrugation height.

## D. Radial Distance of the First Corrugation from the center (plane_rad)

While the corrugated configuration near to coils provides torque increment, the field lines closer to the center of the disc are better utilized by a plain disc configuration due to the configuration of the magnetic field lines. Fibril chain formation occurs optimally when the field lines are directed normally to the transmission surfaces. So, it is seen that by moving the corrugations to the right extreme, i.e., to a larger radial distance, better torque transmission is achieved. The torque values obtained for different values of plane_rad are given in Fig 7. It is observed that the increase in torque transmitted with respect to plane radius has a nonlinear characteristic of varying positive slope, which tends to saturate over a point.

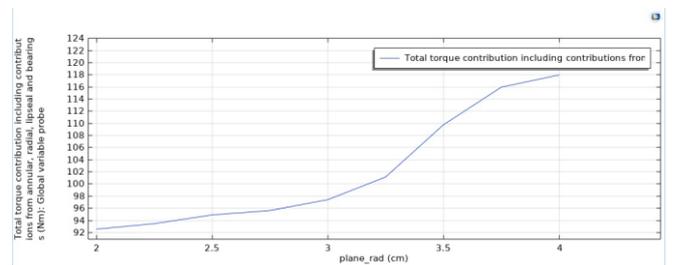

Fig. 7. 2D plot of the variation of the total torque transmitted with change in plane_rad.

## E. Height of individual corrugations (corrugation_height_1, corrugation_height_2, corrugation_height_3)

The effect of the variation in the height of the individual corrugations were studied using a surface plot as depicted in Figure 8. It is of interest to note that the maximum torque transmission occurs when all the three corrugations are of equal height. As the height of one of the corrugations is fixed, the maximum torque transmission capability occurs when the rest of the corrugation heights equals the fixed parameter.

It can be noted that the maximum transmission occurs when the corrugation heights are equal as pointed out by figures 8.(a), 8.(b), 8.(c) and 8.(d). For all these configurations, the maximum torque transmission can be obtained at the combination of equal heights in the surface plots.

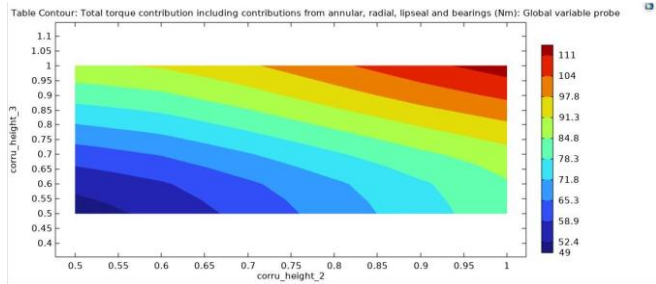
(a.)

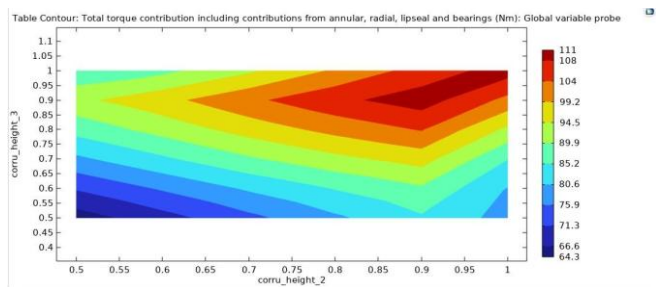
(b.)

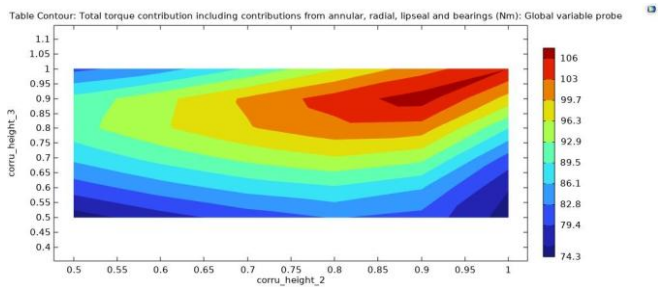
(c.)

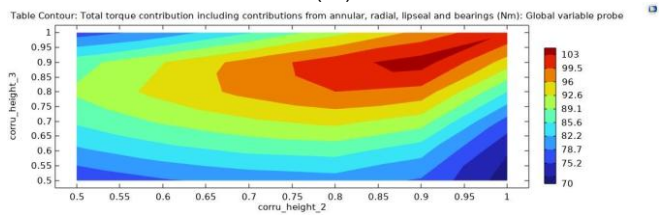
(d.)

Fig. 8. Surface plot of the variation of total torque transmission with respect to variation in the height of individual corrugation 2 and corrugation 3: (a.) when the height of corrugation 1 is 1 cm. (b.) when the height of corrugation 1 is 0.9 cm. (c.) when the height of corrugation 1 is 0.8 cm. (d.) when the height of corrugation 1 is 0.7 cm.

## V. COMPLETE OPTIMIZATION

Optimization for the hybrid corrugated MR clutch was executed using Nelder-Mead optimization technique with the help of the penalty method. Fig.9 depicts the magnetic analysis surface plot of the MR clutch with parameters before the optimization. The optimization was done for the total torque as the objective function with maximization as aim. The total torque transmission capability for the model was 87.06 Nm. Parameters of variation are listed in Table III. Both the initial as well as optimized values are provided for reference. It was observed that for transmission capability to be enhanced, the corrugation height was to be increased and the width of the disc decreased.

Fig. 10 depicts the magnetic analysis surface plot for the optimized geometry of the hybrid corrugated plane type MR Clutch. The achieved torque is 121.34 Nm and is able to accommodate higher transmission capability than the non-optimized geometry.

It is of interest to note that maximum torque transmission occurs when the first corrugation is thicker and the middle corrugation thinner. This configuration is preferred by the optimizing algorithm due to the structures ability to be perpendicular to the magnetic field lines and hence cause maximum transmission capability.

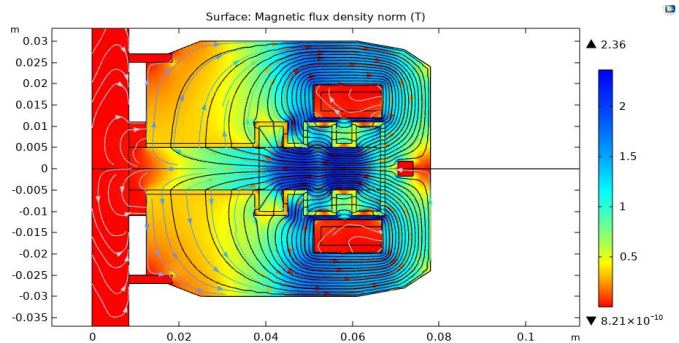

Fig. 9. Surface plot of the magnetic field density along the non-optimized axi-symmetric geometry of the MR Clutch

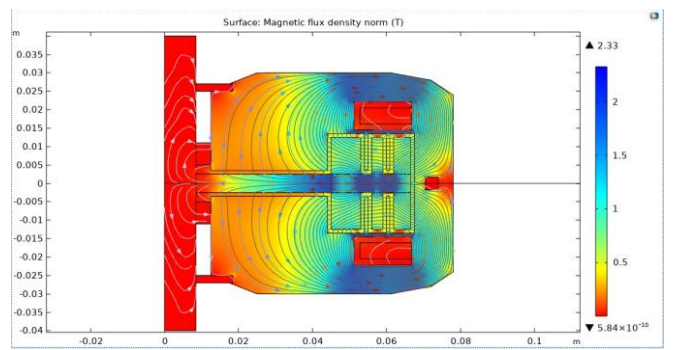

Fig. 10. Surface plot of the magnetic field density along the optimized axi-symmetric geometry of the MR Clutch

TABLE III. THE PARAMETERS OF OPTIMIZATION

| Varied Parameters | Initial value | Optimized value |
|---|---|---|
| bobbin_width | 2 mm | 1.6 mm |
| disc_width | 10 mm | 4.98 mm |
| corrugation_height_1 | 5 mm | 10 mm |
| rugation_height_2 | 5 mm | 10 mm |
| rugation_height_3 | 5 mm | 10 mm |
| plane_rad | 30 mm | 36.5 mm |

## VI. Results and discussions

The hybrid corrugated design of MR clutches was studied using parameters of variations and it was found that the present design is influenced more by variations in the annular gap rather than the axial gaps. Also, it was noted that the bobbin thickness has a decremental impact upon the transmission capability of the MR Clutch.

It was simulated and surmised that the disc width and corrugation height had an optimum combination to yield the maximum transmission capability by a balance of magnetic flux density and transmission surface area. The corrugations when positioned at the maximum radial distance yielded the best transmission capability. Also, the height of the corrugations was altered to yield the optimum torque transmission and it is of interest to note that the heights being equal is the optimum configuration for the corrugations.

The hybrid corrugated plane type MR Clutch design was optimized, and the torque transmission capability was increased from 87.06 Nm to 121.34 Nm. An increase in the transmission capability by 39.37% was attained through this optimization.

TABLE IV. THE PARAMETERS OF OPTIMIZATION

| Characteristic parameter | Un-Optimized Hybrid corrugated plane MR Clutch | Optimized Hybrid corrugated plane MR Clutch |
|---|---|---|
| Torque obtained | 87.06 Nm | 121.34 Nm |
| Average magnetic field strength | 0.522 T | 0.469 T |

## VII. Conclusion

The hybrid corrugated plane type transmission surface for MR clutches has been analytically explored in this study. Also, a penalty method employed optimization was employed in order to achieve an enhanced design with hybrid corrugated plane type transmission surfaces which was found to have a 39.37% augmented transmission capability when compared with its predecessor. Various parameters of interest to the design were identified and a parametric study was employed to capture the effect of these variables upon the clutch performance. The parameters of interest which were studied include the MR gaps consisting of annular and radial surfaces, radii of plane surface, bobbin thickness, disc width and corrugation heights. The optimization was carried out with inclusion of the above-mentioned variables as control variables along with the individual corrugation widths. The optimized model suggests that the corrugation topography be so oriented and structured such that the magnetic field lines are perpendicular for a strong matrix of fibrillations in the MR fluid.